\documentclass[12pt]{article}
\usepackage{amsmath, amsthm, amsfonts}
\usepackage[margin=1 in]{geometry}
\usepackage{blindtext, xcolor}
\usepackage{algorithm}
\usepackage{algorithmicx}
\usepackage{algpseudocode}
\usepackage{apacite}
\usepackage{natbib}
\usepackage{appendix}
\usepackage{rotating}
\usepackage{hyperref}
\usepackage{float}
\usepackage{graphicx}
\usepackage{caption,subcaption}
\usepackage{cleveref}
\usepackage{arydshln}
\usepackage{authblk}
\usepackage{xr,bm}
\usepackage{mathrsfs}
\usepackage{soul} 
\setstcolor{red} 
\usepackage{forest}
\usepackage{booktabs}
\usepackage{multirow}

\newcommand{\be} {\begin{eqnarray*}}
	\newcommand{\ee} {\end{eqnarray*}}

\newcommand{\abs}[1]{\left\vert#1\right\vert}

\theoremstyle{definition}

\newtheorem{theorem}{Theorem}
\newtheorem*{theorem*}{Theorem}

\def\*#1{\bm{#1}}

\title{Generalized Tree-Informed Mixed Model Regression}
\author[1]{Jeremiah Allis}
\author[2]{Xin Jin}
\author[1]{Riddhi Pratim Ghosh}

\affil[1]{Department of Mathematics and Statistics, Bowling Green State University}
\affil[2]{Department of Mathematics, The University of Tampa}
\date{ }

\begin{document}
\maketitle

\begin{abstract}

The standard regression tree method applied to observations within clusters poses both methodological and implementation challenges. Effectively leveraging these data requires methods that account for both individual-level and sample-level effects. We propose Generalized Tree-Informed Mixed Model (GTIMM), which replaces the linear fixed effect in a generalized linear mixed model (GLMM) with the output of a regression tree. Traditional parameter estimation and prediction techniques, such as the expectation-maximization algorithm, scale poorly in high-dimensional settings, creating a computational bottleneck. To address this, we employ a quasi-likelihood framework with stochastic gradient descent for optimized parameter estimation. Additionally, we establish a theoretical bound for the mean squared prediction error. The predictive performance of our method is evaluated through simulations and compared with existing approaches. Finally, we apply our model to predict country-level GDP based on trade, foreign direct investment, unemployment, inflation, and geographic region.

\end{abstract}

\noindent
{\it Keywords:} tree based regression; clustered data; mixed effects; penalized quasi-likelihood; prediction; stochastic gradient descent
\vfill

\section{Introduction}

In the analysis of complex data structures, combining the strengths of tree-based modeling and regression techniques offer a powerful approach for capturing both local and global patterns within a dataset. Decision trees excel at partitioning the predictor space into distinct regions, essentially identifying clusters of similar data points. However, they fall short when it comes to accounting for random effects and identifying unique relationships between predictors and the response variable. Linear mixed models are adept at handling such relationships by being easily interpretable and incorporating random effects, but they can struggle to address the non-linear interactions that decision trees naturally detect. By integrating these two methodologies, we develop a model that leverages the partitioning ability of decision trees to tailor region-specific linear mixed models. This allows for a nuanced and flexible analysis of modeling relationships within clustered data on multiple levels. 

Regression trees have been extensively studied due to their interpretability and effectiveness in capturing non-linear relationships. \cite{breiman1976general} pioneered the combination of decision tree models with linear regression analysis taking an exploratory look into error analysis before \cite{breiman1984classification} developed the CART method to fit constant models on the terminal nodes of a tree. \cite{alexander1996treed} detailed an algorithm that recursively fits data into partitions by calculating the splits of the tree prior to fitting a linear regression. \cite{chaudhuri1994piecewise} proposed an algorithm to recursively use regression trees to fit fixed order piece-wise polynomial functions.  \cite{chaudhuri1995generalized} advanced their work by recursively partitioning data to maximize the likelihood function for generalized regression trees which broadened the application of this model. 
\cite{loh2002regression} introduced the GUIDE algorithm, which addressed variable selection bias and improved interaction detection in regression trees through hypothesis testing. \cite{dusseldorp2010combining} introduced a method, STIMA, that iteratively partitions the predictor space to fit regression models within each node, particularly focusing on optimizing splits and improving prediction accuracy. \cite{dumitrescu2022machine} used many short-depth decision tree models to detect non-linear threshold effects integrated into a penalized logistic regression framework. A limitation of all these models was that they only fit linear (or logistic) regression models and did not address incorporating mixed effects. Mixed effect models allow for the modeling of random effects which account for dependencies and correlation between data points, which is an advantage over just fixed effect models \cite{zimmerman2020linear}. 

\cite{sela2012re} developed the RE-EM tree, which alternates estimating the fixed and random effects until convergence for a mixed model regression tree. \cite{loh2013regression} extended GUIDE to longitudinal and multivariate data and focused on variable selection through multiple-step interaction detection. \cite{eo2014tree} estimated their model by minimizing residual-based node impurity to optimize the tree fit. \cite{fokkema2018detecting} presented a GLMM tree focused on modeling treatment subgroups in medical research when handling clustered data. All of these approaches do not focus on a likelihood based approach, but rather getting focus only on variable selection and estimation through a recursive partitioning algorithm. \cite{hajjem2011} introduced a mixed-effects regression tree model that can handle unbalanced clusters through an expectation-maximization, likelihood approach. \cite{hajjem2017generalized} expanded on this approach by generalizing the model using a penalized quasi-likelihood approach.  While these approaches do include a quasi-likelihood based approach, they use an expectation maximization (EM) approach to optimization which does not scale well to large data sets as discussed by \cite{chen2018stochastic}.

\cite{chipman2010bart} proposed a Bayesian approach to this problem and backfitting a sum-of-trees model with a Markov Chain Monte Carlo algorithm to explain the variation in the model using multiple decision trees. Recently, \cite{linero2024generalized} improved on this method so that conjugate priors are not needed in the modeling process. Another evolution of this modeling problem was to use generalized random forest. \cite{atheygrf} introduced generalized random forests by exploring their asymptotic properties when performing various statistical tasks. \cite{capitaine2021random} extended generalized random forests to consider high-dimensional, longitudinal data and better models the covariance structures when compared to standard random forests. All of these models are extensions of tree-based regression, but apply methods that are outside the scope of this paper. 

We further explore tree-based regression by using a quasi-likelihood based approach with a focus on parameter estimation with a stochastic gradient descent algorithm. \cite{breslow1993approximate} applied this approach with traditional linear mixed models, and more recently, \cite{mandel2023neural} expanded this approach by using a neural network to model the fixed effect in a nonlinear setting. In this paper, each node of a decision tree will represent one region of data in the predictor space, a regression model will be fit over each region, and a global random effect will be included to capture any lingering relationships spanning the entire dataset. The decision trees will be fit using methods described by \cite{james2013introduction}, but the method described in this paper will focus on maximizing the quasi-likelihood function rather than determining the best splits of the decision tree. Other methods used will be stochastic gradient descent (SGD) and best linear unbiased predictors (BLUP) to handle the random effects. By incorporating this approach, the model aims to achieve both precise local regression fits within tree-defined clusters and capture the correlation of the entire data set through random effects, setting the stage for a more comprehensive exploration of tree-based regression techniques. 

The remainder of this paper is organized as follows: \Cref{model_notation} presents the proposed model and notation. \Cref{methodology} details the methodology for parameter estimation. In \Cref{simulation}, a simulation study is conducted to compare the predictive performance of the proposed model against traditional methods followed by an analysis of country data, predicting GDP from trade, foreign direct investments, and
unemployment metrics, in \Cref{dataanalysis}. Finally, \Cref{discussion} concludes this article with a discussion. 

\section{Model and Notation} \label{model_notation}

The linear mixed model (LMM) is designed to analyze data that exhibit both fixed and random effects. Suppose there are $N$ observations, where each observation $Y_i$ depends on a set of predictors $\bm{X}_i=(X_{i1}, \ldots, X_{ip})^\top$ corresponding to the fixed effects and $\bm{Z}_i=(Z_{i1}, \ldots, Z_{iq})^\top$ corresponding to the random effects. The fixed effects $\bm{\beta}=(\beta_1, \ldots, \beta_p)^\top$ are the regression parameters in the linear model, while the random effects $\bm{b}=(b_1, \ldots, b_q)^\top$ account for variations specific to different clusters or groups within the data. Assuming the random noise is represented by $\bm{\varepsilon}=(\varepsilon_1, \ldots, \varepsilon_N)^\top$, 
the traditional LMM can be written as
\begin{equation}
    Y_i = \bm{X}_i^\top\bm{\beta} + \bm{Z}_i^\top\bm{b} + \varepsilon_i,
    \nonumber
\end{equation}
where $\epsilon$ is assumed to follow a normal distribution. 
In this model, the fixed effect is determined by a linear regression, $\bm{X}_i^\top\bm{\beta}$. In our model, the fixed effect is replaced by a linear regression to each terminal node of a decision tree accounting for a clustered data structure, such as modeling patient level data across multiple hospitals or predicting crop yield across different plots of land.  


\begin{figure}[h!]
\centering
\begin{minipage}{0.45\textwidth} 
\centering
\scalebox{0.8}{
\begin{forest}
for tree={
  grow=south,
  parent anchor=south,
  child anchor=north,
  edge={-latex},
  rounded corners,
  draw,
  align=center,
  l sep+=15pt,
  s sep+=15pt
}
[$X_1 \leq t_1$
  [$X_2 \leq t_2$
    [$R_1$]
    [$X_1 \leq t_3$
      [$R_2$]
      [$R_3$]
    ]
  ]
  [$X_1 \leq t_4$
    [$R_4$]
    [$X_2 \leq t_5$
      [$R_5$]
      [$R_6$]
    ]
  ]
]
\end{forest}
}
\end{minipage}%
\begin{minipage}{0.45\textwidth}
\centering
\scalebox{0.8}{
\begin{tikzpicture}

    \draw[thick] (0,0) rectangle (6,6);

    \draw[thick] (2,0) -- (2,6);  
    \draw[thick] (1,2) -- (1,6);  
    \draw[thick] (4,0) -- (4,6);  
    \draw[thick] (0,2) -- (2,2);  
    \draw[thick] (4,4) -- (6,4);  

    \node at (1, 1) {$R_1$};  
    \node at (0.5, 4) {$R_2$};  
    \node at (1.5, 4) {$R_3$};   
    \node at (3, 3) {$R_4$}; 
    \node at (5, 1) {$R_5$};   
    \node at (5, 5) {$R_6$};  

    \node at (2,-0.3) {$t_1$};  
    \node at (4,-0.3) {$t_4$}; 
    \node at (-0.3,2) {$t_2$};
    \node at (1,6.3) {$t_3$}; 
    \node at (6.3,4) {$t_5$};   

    \node at (3, -1) {\large $X_1$};
    \node at (-1, 3) {\large $X_2$};
    
\end{tikzpicture}
}
\end{minipage}
\captionsetup{labelfont=bf}
\caption{\textbf{Representation of a Decision Tree.} On the left is a decision tree model partitioning the data on two predictors $X_1$ and $X_2$ at splits $t_1, \ldots, t_5$ into regions $R_1, \ldots, R_6$. On the right is a two-dimensional predictor space split up into 6 regions based on the tree structure on the right.}
\label{modfig1}
\end{figure}
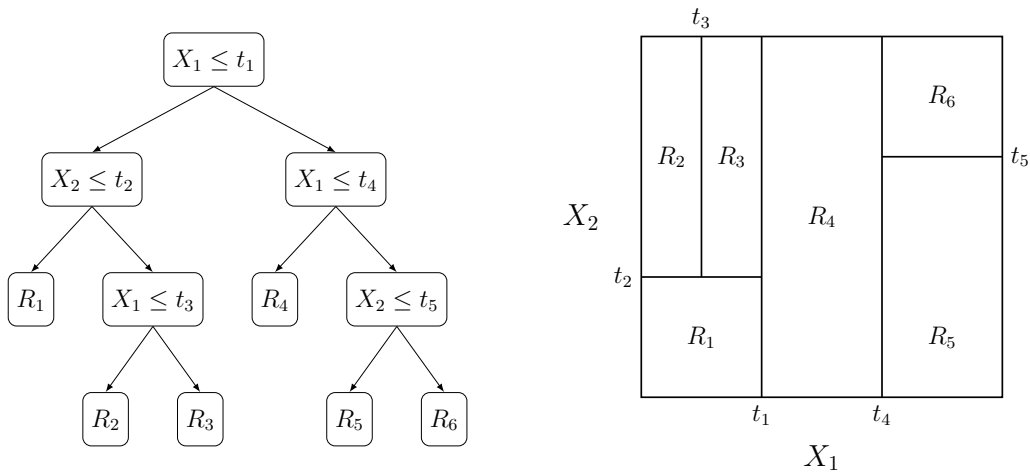

The tree structure splits the data into subsets based on the values of predictors, where each internal node represents a decision, each branch represents the outcome of the decision, and each leaf node represents one subset of our data. The output of a decision tree is based on minimizing an objective function, but the specific predictions of the tree are not important for our model; we care only about the number of terminal nodes and  grouping of the data. The subgroups created by the tree will indicate which data are similar enough to train the same fixed-effect model.\Cref{modfig1} shows a tree with six leaf nodes and how it splits the predictor space into six distinct regions. 

In the Generalized Tree Informed Mixed Model (GTIMM), we represent regions in the data by considering indicator variables $\bm{R}_i=(R_{i1}, \ldots, R_{iM})^\top$ that represent which node of the tree (or region of the predictor space) an observation falls into. For any observation, the sum of the indicator is one since it falls into only one region. There are $n_1, \ldots, n_M$ observations in the $M$ regions where $\sum_{m=1}^M n_m = N$. We assume that $M \ll N$. Let $\bm{\beta}^* = \left(\bm{\beta}^{(1)}, \ldots, \bm{\beta}^{(M)} \right)$, where $\bm{\beta}^{(m)} =\left(\beta^{(m)}_1, \ldots, \beta^{(m)}_p \right)^\top$ represents the  coefficients for $m^{th}$ region. The remainder of the model specifications have already been defined in the LMM case. For predicting \( Y_i\), the GTIMM is modeled as

\begin{equation} \label{modeq1}
    Y_i = \bm{X}_i^\top \bm{\beta}^* \bm{R}_i + \bm{Z}_i^\top\bm{b}+\varepsilon_i.
\end{equation}

Alternatively, we can define the regions such that $\bm{R} = (\bm{R}_1, \ldots, \bm{R}_N)$ is an indicator matrix of dimension $M \times N$ and let $\bm{1}_M$ be a vector of ones of length $M$. Then, the model in \Cref{modeq1} can be adjusted for predicting a vector \( \bm{Y} \), 
\begin{equation} \label{modeq2}
    \bm{Y} = \left((\bm{X \beta}^*) \odot \bm{R}^{\top} \right) \bm{1}_M + \bm{Z}\bm{b}+\bm{\varepsilon}, 
\end{equation}
where $\odot$ denotes the element-wise multiplication between matrices. \\

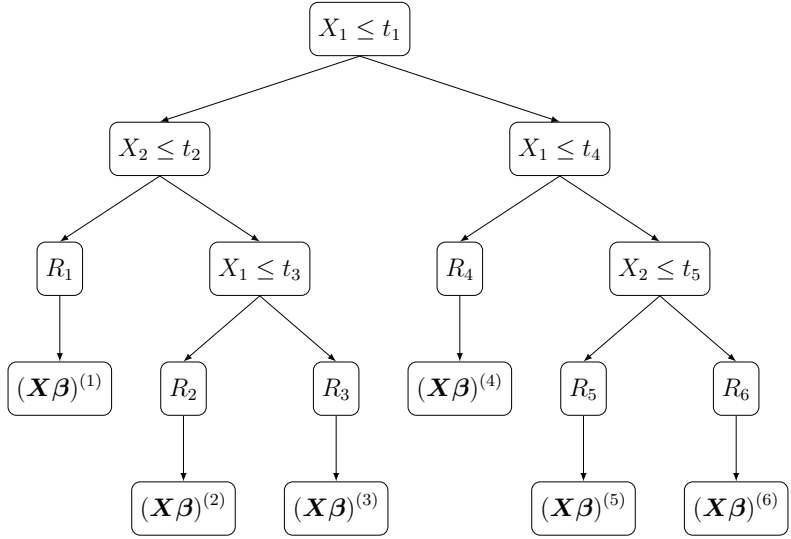
\begin{figure}[h!]
\centering
\scalebox{0.8}{
\begin{forest}
for tree={
  grow=south,
  parent anchor=south,
  child anchor=north,
  edge={-latex},
  rounded corners,
  draw,
  align=center,
  l sep+=15pt,
  s sep+=15pt
}
[$X_1 \leq t_1$
  [$X_2 \leq t_2$
    [$R_1$
      [{$\displaystyle (\bm{X} \bm{\beta})^{(1)} $}]
    ]
    [$X_1 \leq t_3$
      [$R_2$
        [{$\displaystyle (\bm{X} \bm{\beta})^{(2)} $}]
      ]
      [$R_3$
        [{$\displaystyle (\bm{X} \bm{\beta})^{(3)} $}]
      ]
    ]
  ]
  [$X_1 \leq t_4$
    [$R_4$
      [{$\displaystyle (\bm{X} \bm{\beta})^{(4)} $}]
    ]
    [$X_2 \leq t_5$
      [$R_5$
        [{$\displaystyle (\bm{X} \bm{\beta})^{(5)} $}]
      ]
      [$R_6$
        [{$\displaystyle (\bm{X} \bm{\beta})^{(6)} $}]
      ]
    ]
  ]
]
\end{forest}
}
\captionsetup{labelfont=bf}
\caption{\textbf{Representation of GTIMM Fixed Effects.} An extension of the decision tree in \Cref{modfig1}. The fixed effects for the \( m^{th} \) region are represented by \( (\bm{X} \bm{\beta})^{(m)} \), which are $N \times 1$ vectors such that $\sum_{m=1}^{M} (\bm{X}\bm{\beta})^{(m)} = \left((\bm{X \beta}^*) \odot \bm{R}^{\top} \right) \bm{1}_M$ and  \( (\bm{X} \bm{\beta})^{(m)}_i = 0 \) if \( \bm{X}_i \notin R_m\). } 
\label{modfig2}
\end{figure}

While the random effect models the correlation across the entire data set, \Cref{modfig2} illustrates how the fixed effects are handled by multiple linear models—one for each region identified by the tree splits. This framework enables the model to capture distinct subgroup-specific trends, ensuring that genuinely different subpopulations obtain separately estimated linear models. This yields more nuanced inferences about the overall population than what a decision tree typically offers. 


\section{Methodology} \label {methodology}

We now describe how parameters are estimated in the proposed model and how we assess prediction error. 
Section \ref{parameter_estimation} outlines our quasi-likelihood approach with BLUP for random effects and how SGD is used in the estimation. 
Section \ref{error_estimation} discusses estimating and bounding predictive error, including a theoretical 
comparison of GTIMM with a traditional GLMM.

\subsection{Parameter Estimation}
\label{parameter_estimation}
Our approach to estimation for the coefficients for $\bm{\beta}^*$ closely follows the approach of \cite{breslow1993approximate} and \cite{mandel2023neural} who performed similar analyses on the LMM and neural network mixed models respectively. The parameters are estimated via a quasi-likelihood approach while the random effect vector $\bm{b}$ is estimated using the classical BLUP approach. For a single observation, let the conditional mean and variance of $Y_{i}$ given $\bm{b}$ be 
\begin{center}
    $E \left(Y_{i}|\bm{b} \right) = \mu_{i}^{\bm{b},(m)}$,\\
    $\text{Var}\left(Y_{i}|\bm{b} \right) = \phi \alpha_{im} v \left(\mu_{i}^{\bm{b},(m)} \right)$,
\end{center}
where $\mu_{i}^{\bm{b},(m)}$ is a function of $\bm{\beta}$, $v(\cdot)$ is a known variance function, $\alpha_{im}$ is a known constant, and $\phi$ is a dispersion parameter. The link function relating the conditional mean to the linear predictor $\eta_{i}^{\bm{b},(m)} = \sum_{m=1}^M {\bm{X}_i^{(m)}}^\top \bm{\beta}^{(m)} + \bm{Z}_i^\top\bm{b}$ will be $g \left(\mu_{i}^{\bm{b},(m)} \right) = \eta_{i}^{\bm{b},(m)}$ with an inverse defined as $h=g^{-1}$. We assume that the covariance matrix $\bm{\Sigma}_{b}$ is dependent on an unknown vector $\bm{\theta}$ of variance components, so the integrated quasi-likelihood equation is written as 

\begin{equation}\label{quasi.integral}
    \exp\{ql(\bm{\beta}^*, \bm{\theta})\} \propto \abs{\bm{\Sigma}_{b}}^{-\frac{1}{2}} \int \exp \left\{ \frac{1}{\phi} \sum_{m=1}^M \sum_{i=1}^{n_m} \int_{Y_{i}}^{\mu_{i}^{\bm{b},(m)}} \frac{Y_{i}-u}{\alpha_{im} v(u)} du - \frac{1}{2}\bm{b}^{\top} \bm{\Sigma}_{b}^{-1}\bm{b} \right\}  d\bm{b}.
\end{equation}

Since this integral is difficult to evaluate, we follow \cite{barndorff1979edgeworth} by using a Laplace approximation of the outside integral by writing \Cref{quasi.integral} in the form of 
\begin{center}
    $\displaystyle c|\bm{\Sigma}_{b}|^{-\frac{1}{2}} \int \exp\{-\kappa(\bm{b})\} d\bm{b}$, 
\end{center}

where

\begin{center}
    $ \kappa(\bm{b}) = \displaystyle - \frac{1}{\phi} \sum_{m=1}^M \sum_{i=1}^{n_m} \int_{Y_{i}}^{\mu_{i}^{\bm{b},(m)}} \frac{Y_{i}-u}{\alpha_{im} v(u)} du + \frac{1}{2}\bm{b}^{\top} \bm{\Sigma}_{b}^{-1}\bm{b}$.
\end{center}
Let $\kappa'$ and $\kappa''$ represent the $q$ length vector and $q \times q$ matrix of the first and second-order partial derivatives of $\kappa$. Then, the Laplace approximation yields the following result:
\begin{equation}
    ql(\bm{\beta}^*, \bm{\theta}) \approx -\frac{1}{2}\log{\abs{\bm{\Sigma}_{b}}}-\frac{1}{2}\log{|\kappa''(\tilde{\bm{b}}) |} - \kappa'(\tilde{\bm{b}}),  \nonumber
\end{equation}
where $\tilde{\bm{b}}$ is the mode of $\kappa(\bm{b})$ obtained by solving 
\begin{equation} 
    \kappa'(\bm{b}) = - \frac{1}{\phi} \sum_{m=1}^M \sum_{i=1}^{n_m} \frac{\left(Y_{i}-\mu_{i}^{\bm{b},(m)} \right) \bm{Z}_{i}}{\alpha_{im} v \left(\mu_{i}^{\bm{b},(m)} \right) g' \left(\mu_{i}^{\bm{b},(m)} \right)} + \bm{\Sigma}_{b}^{-1}\bm{b} = \bm{0}. \nonumber
\end{equation}

The second-order derivative yields $\kappa''(\bm{b}) = \bm{Z}^{\top}\bm{WZ} + \bm{\Sigma}_{b}^{-1} +\bm{V} \approx \bm{Z}^{\top}\bm{WZ} + \bm{\Sigma}_{b}^{-1} $ where $\bm{W}$ is a block diagonal matrix with the $M$ blocks where the $m^{th}$ block is $W^{(m)}$ which is a $n_m \times n_m$ diagonal matrix with terms
\begin{center}
    $\left(W^{(m)} \right)_{ii} = \left[\phi \alpha_{im} v \left(\mu_{i}^{\bm{b},(m)} \right) \left\{ g''(\mu_{im}^{\bm{b}}) \right\}^2 \right]^{-1}$, 
\end{center}
for $i=1, \ldots, n_m$, and the remainder term 
\begin{center}
    $\bm{V} = \displaystyle - \frac{1}{\phi} \sum_{m=1}^M \sum_{i=1}^{n_m} \left(Y_{i} -\mu_{i}^{\bm{b},(m)} \right) \bm{Z}_{i} \frac{\partial}{\partial \bm{b}} \left[\frac{1}{\alpha_{im} v \left(\mu_{i}^{\bm{b},(m)} \right)g' \left(\mu_{i}^{\bm{b},(m)} \right)} \right]$
\end{center}
is higher than second order and has an expectation of $\bm{0}$. The resulting approximation can be written as 
\begin{equation}\label{quasi.intergral.approx}
    \exp(ql\{\bm{\beta}^*, \bm{\theta}\}) \propto \abs{\bm{\Sigma}_{b}}^{-\frac{1}{2}}\abs{ \bm{Z}^{\top}\bm{WZ} + \bm{\Sigma}_{b}^{-1} }^{-\frac{1}{2}} \exp \biggl\{ \frac{1}{\phi} \sum_{m=1}^M \sum_{i=1}^{n_m} \int_{Y_{i}}^{\mu^{\tilde{\bm{b}}, (m)}_{i}} \frac{Y_{i}-u}{\alpha_{im} v(u)} du - \frac{1}{2}\tilde{\bm{b}}^{\top}\bm{\Sigma}_{b}^{-1}\tilde{\bm{b}} \biggl\}. 
    \nonumber
\end{equation}

Assuming that the GLM weights vary slowly, or not at all, as a function of $\mu^{\bm{b}, (m)}_{i}$, then $\bm{W}$ is negligible and can be ignored. The resulting approximated log quasi-likelihood is
\begin{equation}\label{quasi.approx} 
    ql(\bm{\beta}^*, \bm{\theta}) \propto \frac{1}{\phi} \sum_{m=1}^M \sum_{i=1}^{n_m} \int_{Y_{i}}^{\mu^{\tilde{\bm{b}}, (m)}_{i}} \frac{Y_{i}-u}{\alpha_{im} v(u)} du - \frac{1}{2}\tilde{\bm{b}}^{\top}\bm{\Sigma}_{b}^{-1}\tilde{\bm{b}},
    \nonumber
\end{equation}
where \(\tilde{\bm{b}}\) is obtained by maximization, and partial derivatives with respect to each $\bm{\beta}^{(m)}$ are given by

\begin{equation}
    \frac{\partial ql}{\partial \bm \beta^{(m)}} = \frac{1}{\phi} \sum_{i=1}^{n_m} \frac{\left(Y_{i}-\mu^{\tilde{\bm{b}},(m)}_{i} \right)}{\alpha_{im} v \left(\mu_{i}^{\tilde{\bm{b}},(m)} \right) g' \left(\mu_{i}^{\tilde{\bm{b}},(m)} \right)} \bm{X}_i^{(m)}. \nonumber
\end{equation}

Setting $\displaystyle \frac{\partial ql}{\partial \bm \beta^{(m)}} = \bm{0}$, we obtain the estimators $\hat{\bm{\beta}}^{(m)}$. The estimates are found in a computationally efficient manner via stochastic gradient descent, which maximizes the quasi-likelihood equations.

One of the first derivations of the BLUP method was with a linear mixed model by \cite{henderson1963selection}. The BLUP is an estimate for a random effect such that it minimizes the mean squared error and is unbiased. The method requires to assume a joint distribution relationship between $\bm{Y}$ and $\bm{b}$ and then apply an approach similar to maximum likelihood estimation; take the derivative of the log-likelihood of the joint distribution and find it's maximum. For a mixed model of the form 
\begin{center}
    $\bm{Y} = \bm{X \beta} + \bm{Z}\bm{b}+\bm{\varepsilon}$,  
\end{center}
where the variance matrices and distributions are defined in \Cref{model_notation}, it is found that
\begin{center}
    $\hat{\bm{b}} = \bm{\Sigma_{b}} \bm{Z}^{\top} (\bm{\Sigma_{\varepsilon}} + \bm{Z}\bm{\Sigma_{b}} \bm{Z}^{\top})^{-1} \left(\bm{Y} -\bm{X} \hat{\bm{\beta}} \right)$, 
\end{center}
see \cite{henderson1975best} for more details. 
The GTIMM takes a similar form, thus the BLUP estimate will be similar. Assume the joint distribution of $\bm{Y}$ and $\bm{b}$ is 
\begin{center}
    $(\bm{Y}, \bm{b})^\top \sim \text{MVN} \left( \begin{pmatrix} \left((\bm{X \beta}^*) \odot \bm{R}^{\top} \right) \bm{1}_M \\
    \bm{0}
    \end{pmatrix}
    , 
    \begin{pmatrix} 
    \bm{\Sigma_\varepsilon} + \bm{Z}\bm{\Sigma_{b}}\bm{Z}^{\top} & \bm{Z}\bm{\Sigma_{b}} \\ \bm{\Sigma_{b}}\bm{Z}^{\top} & \bm{\Sigma_{b}} 
    \end{pmatrix} \right)$. 
\end{center}
The log-likelihood of the joint distribution is 
\begin{align*}
    \begin{split}
    \ell(\bm{\beta}^*, \bm{b}) \propto
    & -\frac{1}{2} \left\{\left(\bm{Y} - \left((\bm{X \beta}^*) \odot \bm{R}^{\top} \right) \bm{1}_M - \bm{Z}^{\top}\bm{b} \right)^{\top} (\bm{\Sigma_{\varepsilon}} + \bm{Z}\bm{\Sigma_{b}} \bm{Z}^{\top})^{-1} \right.\\
    & \hspace{3cm} \left. \left(\bm{Y} - \left((\bm{X \beta}^*) \odot \bm{R}^{\top} \right) \bm{1}_M -\bm{Z}^{\top}\bm{b} \right) - \bm{b}^{\top} \bm{\Sigma_b}^{-1} \bm{b} \right\}.
    \end{split}
\end{align*}
Taking the derivative and solving for $\bm{b}$ will show that
\begin{equation}
    \hat{\bm{b}} = \bm{\Sigma_{b}} \bm{Z}^{\top} (\bm{\Sigma_{\varepsilon}} + \bm{Z}\bm{\Sigma_{b}} \bm{Z}^{\top})^{-1} \left(\bm{Y} - \left((\bm{X} \hat{\bm{\beta}}^*) \odot \bm{R}^{\top} \right) \bm{1}_M \right). 
    \nonumber
\end{equation} 

From there, we obtain a closed-form estimate for the random effect coefficients. By applying BLUP, we can separate the fixed effects from the random effects, which allows for more accurate modeling of the clustering structure of the data. The BLUP approach minimizes the mean squared error and provides the best prediction of the random effects, taking into account both the fixed effects and the variability within the random effects. It assumes that the random effects are normally distributed and uses the variance-covariance structure of the random effects to improve predictions.

We find our model’s optimal parameters by applying stochastic gradient descent. We iteratively sample batches of observations, compute the gradient of the quasi-likelihood with respect to the region-specific coefficients and random effects, then update those parameters incrementally. This process continues until the parameters converge, which happens when the quasi-likelihood function is optimized. The SGD updates occur on smaller subsets of the data rather than the entire dataset at once, so the method remains both computationally efficient and capable of scaling to larger problems.


\subsection{Error Estimation}
\label{error_estimation}

The MSPE for GTIMM and GLMM are
\begin{center}          
    $\text{MSPE}_{\widehat{\text{GTIMM}}} = E \left[Y_{i} - g^{-1}\left(\hat{T}_{i}^{(m)} \right) \right]^2$ \\
    $\text{MSPE}_{\widehat{\text{GLMM}}} = E \left[Y_{i} - g^{-1}\left(\hat{M}_{i} \right) \right]^2$,
\end{center}
where $\hat{T}_{i}^{(m)} = \bm{X}_i^\top \hat{\bm{\beta}}^* \bm{R}_i + \bm{Z}_i^\top \hat{\bm{b}}$ and $\hat{M}_{i} = \bm{X}_i^\top \hat{\bm{\beta}} + \bm{Z}_i^\top \tilde{\bm{b}}$ are the GTIMM and GLMM, with $\hat{\bm{b}}$ and $\tilde{\bm{b}}$ being the BLUP and least squares estimator of the random effect vector $\bm{b}$, and $\hat{\bm{\beta}}$ is the least squares estimator of the true regression coefficient vector $\bm{\beta}$. We present a bound on the difference in MSPE between GTIMM and GLMM in the following theorem.

\begin{theorem} \label{thm1}
    Assume a tree-based method is used to estimate the fixed effect in the model
    $Y_i = \bm{X}_i^\top \bm{\beta}^* \bm{R}_i + \bm{Z}_i^\top \bm{b} + \varepsilon_i$, where  $\bm{\beta}^* \in \mathbb{R}^{p \times M}$, $\bm{R}_i \in \mathbb{R}^M$, and $(\bm{b}, \bm{\varepsilon})^\top \sim N(\bm{0}_{2 \times 1}, \text{diag}(\bm{\Sigma_b}, \bm{\Sigma_\varepsilon}))$, with $\bm{b}$ and $\bm{\varepsilon}$ being independent. Furthermore, assume that there exist constants $c_1, c_2 > 0$ such that for all $1 \leq i \leq N$, $\|\bm{X}_i\|_{\infty} \leq c_1 \quad \text{and} \quad \|\bm{Z}_i\|_{\infty} \leq c_2$. Then the difference in MSPE satisfies
    \begin{center}
        $\left|\text{MSPE}_{\widehat{\text{GTIMM}}} - \text{MSPE}_{\widehat{\text{GLMM}}} \right| = \mathcal{O}\left(\frac{M}{N}\right)$.   
    \end{center}
\end{theorem}
The proof of the theorem is deferred to \Cref{appendix}. Both the complexity of the model and the sample size influence predictive performance. As the sample size $N$ increases, the difference between the MSPEs of the two models decreases, and the rate of this decay depends on the ratio of the number of regions $M$ to the sample size $N$. Since $M$ is small relative to $N$, the models are expected to perform similarly as $N$ grows, with the difference in MSPE shrinking at a rate of $\mathcal{O}\left(\frac{M}{N}\right)$. 


\section{Simulations}\label {simulation}

In this section, we compare the performance of our GTIMM model with the linear mixed effects model, a singular decision tree, a random forest, and the GLMM tree from \cite{fokkema2018detecting} by comparing mean squared prediction error. The GLMM tree, or GLMER, is fitted using an r function called \texttt{glmertree}. They fit the GTIMM model using recursion, alternately estimating the fixed effects with a random effect offset and then estimating the random effects using an E-M algorithm. 

\begin{figure}[h!] 
    \centering
    \includegraphics[scale=0.6]{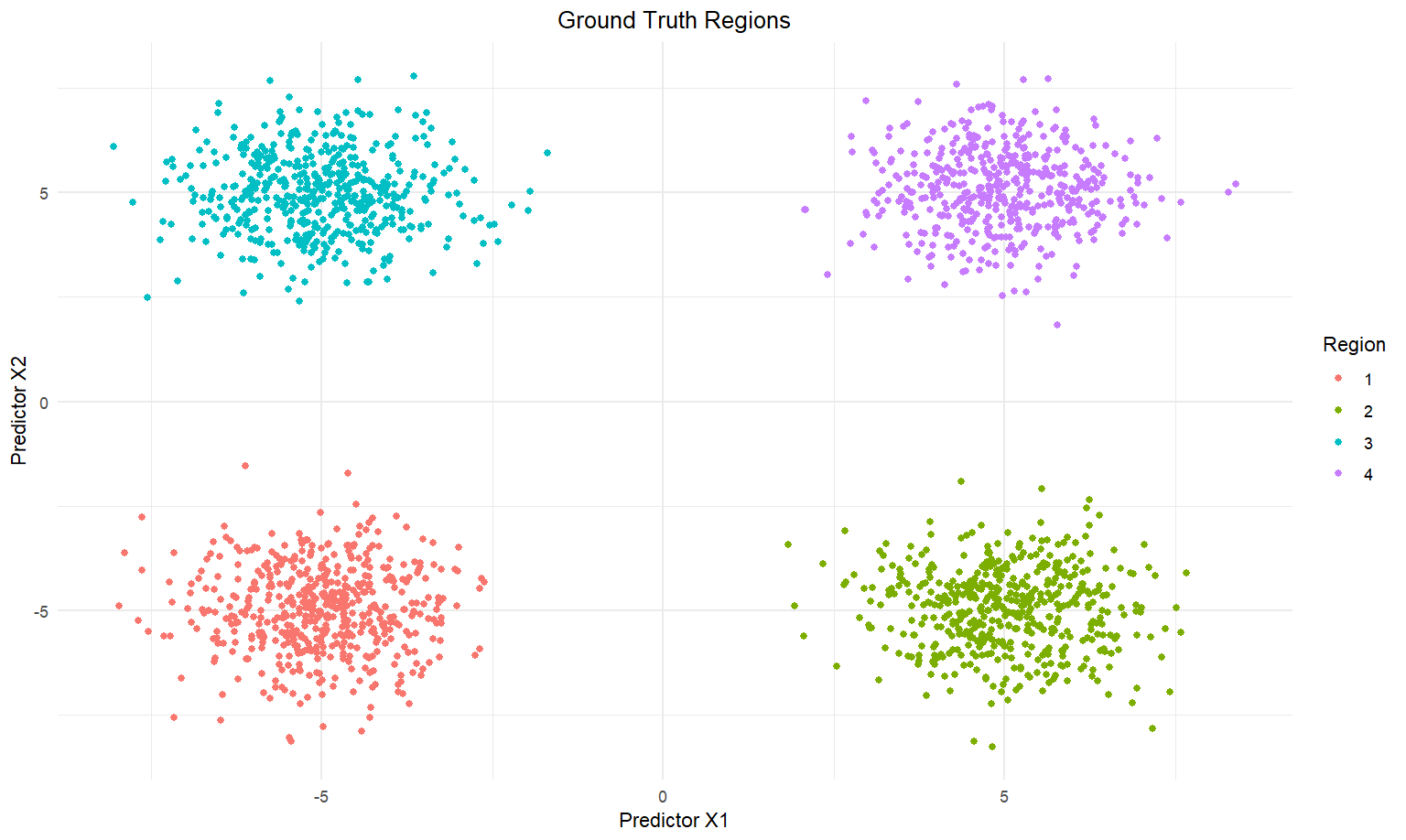}
    \captionsetup{labelfont=bf}
    \caption{ \textbf{Simulated Regions.} Ground truth regions for the simulated data.}\
    \label{simulfig1}
\end{figure}

We consider two covariates, $X_1$ and $X_2$, simulated to ensure well-separated regions as shown in \Cref{simulfig1}. The means of the normal distributions are either 5 or -5, and all have a standard deviation of one. Following this style, 2000 data points, \((X_1, X_2)\), are simulated, evenly distributed across the four regions. As a result, we expect the RF, decision tree, and GTIMM model to perform better than the LMM model which does not take this structure into account.

\begin{equation}\label{simuleq1}
  Y_i = \begin{cases} 
      2+1.5x_1+0.5x_2+\bm{Z}_i^{\top}\bm{b}+\varepsilon_i, & \bm{X}_i\in R_1 \\
      -1+2.5x_1-0.5x_2+\bm{Z}_i^{\top}\bm{b}+\varepsilon_i, & \bm{X}_i\in R_2 \\
      1+-2x_1+1x_2+\bm{Z}_i^{\top}\bm{b}+\varepsilon_i, & \bm{X}_i\in R_3 \\
      -2-1.5x_1-1x_2+\bm{Z}_i^{\top}\bm{b}+\varepsilon_i, & \bm{X}_i\in R_4
   \end{cases}
\end{equation}

The fixed effect coefficients are region-dependent, as seen in equation \eqref{simuleq1}. We consider 10 groups, specified by the coefficients $\bm{b}$ generated from an \(N(0, 2)\) distribution, to model the random effects.The random noise was generated from a standard normal distribution. This simulation provides the hierarchical structure and the partitions necessary for the GTIMM.

\begin{figure}[h!] 
    \centering
    \includegraphics[scale=0.6]{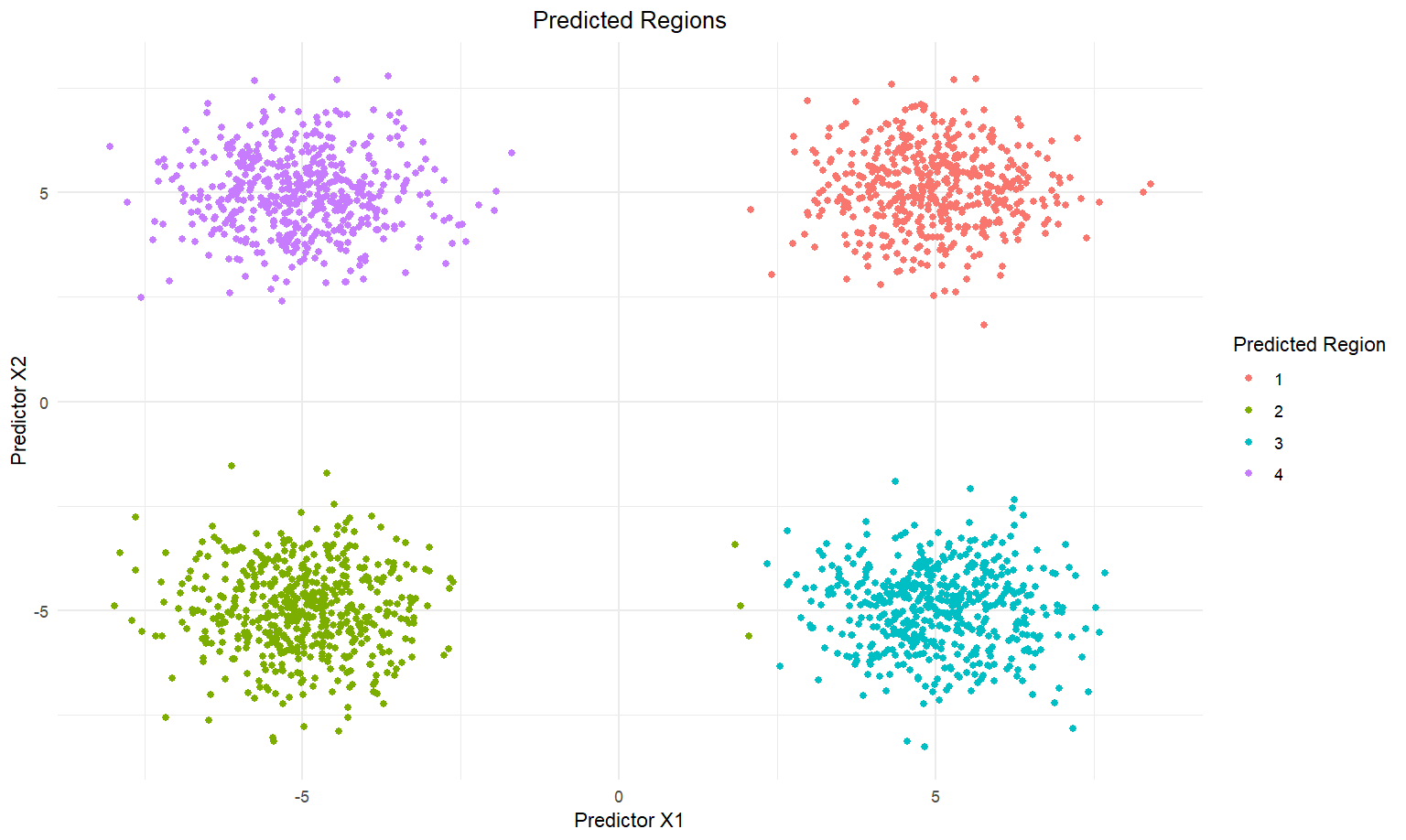}
    \captionsetup{labelfont=bf}
    \caption{\textbf{Decision Tree Regions.} Regions generated by a decision tree forced to produce 4 nodes, when using the whole data set. The regions are similar to the ground truth regions. The tree incorrectly predicted 3 data points. }\
    \label{simulfig2}
\end{figure}

To model the data according to the GTIMM model, we first had to determine the number of terminal nodes. Through 5-fold cross-validation, 4 terminal nodes are obtained, which matches how the data was originally fit. Then, the entire data set is first modeled by a decision tree to gain information about the region-wise indicator variables. \Cref{simulfig2} displays that the region indicator variables capture the simulated regions quite well outside of three data points being grouped into Region 2 instead of Region 3.

\begin{table}[h!] 
    \centering
    \captionsetup{labelfont=bf}
    \caption{\textbf{Fixed Effect Estimation Comparison.} Comparison between the true values of the fixed effect coefficients to the estimated values when using the GTIMM model.}
    \label{simultab1}
    \small
    \begin{tabular}{lccc}
        \toprule
        \textbf{Region} & \textbf{Parameter} & \textbf{True Value} & \textbf{Estimated Value} \\ 
        \midrule
        \multirow{3}{*}{1} & $\beta_0^1$ & 2.0 & 2.107 \\
                            & $\beta_1^1$ & 1.5 & 1.569 \\
                            & $\beta_2^1$ & 0.5 & 0.444 \\ 
        \midrule
        \multirow{3}{*}{2} & $\beta_0^2$ & -1.0 & -1.117 \\
                            & $\beta_1^2$ & 2.5 &  2.622\\
                            & $\beta_2^2$ & -0.5 & -0.456 \\
        \midrule
        \multirow{3}{*}{3} & $\beta_0^3$ & 1.0 & 1.068 \\
                            & $\beta_1^3$ & -2.0 & -1.962 \\
                            & $\beta_2^3$ & 1.0 & 0.9799 \\
        \midrule
        \multirow{3}{*}{4} & $\beta_0^4$ & -2.0 & -1.921 \\
                            & $\beta_1^4$ & -1.5 & -1.476 \\
                            & $\beta_2^4$ & -1.0 & -1.003 \\
        \bottomrule
    \end{tabular}
\end{table}

\Cref{simultab1} compares the results of the estimated region coefficients compared to the values in Equation \eqref{simuleq1} demonstrating consistent estimates of the model parameters.

\begin{table}[h]
    \centering
    \captionsetup{labelfont=bf}
    \caption{\textbf{MSPE Comparison.} Comparison of MSPE for the generalized tree-informed mixed model, the linear mixed model, random forest model, and a singular decision tree.}
    \label{simultab3}
    \small
    \begin{tabular}{lccccc}
        \toprule
        \textbf{Model} & \textbf{GTIMM} & \textbf{LMM} & \textbf{RF} & \textbf{Tree} & \textbf{GLMER} \\ 
        \midrule
        \textbf{MSPE} & 1.367 & 181.075 & 5.322 & 8.024 & 3.054 \\
        \bottomrule
    \end{tabular}
\end{table}

We further compare the predictive performance of GTIMM with other approaches with respect to the mean squared prediction error (MSPE).The values of  MSPE of the naive linear mixed model, decision tree, random forest, and GTIMM models are in \Cref{simultab3}. Each model is fit to the same training data and the test set is used for predictions. This shows clearly that the GTIMM outperforms others demonstrating its robustness in handling clustered data structure. The next lowest value of MSPE is achieved by GLMER model since it is fitting a tree-based regression model using a recursive algorithm.


\section{Data Analysis}\label {dataanalysis}

We analyze a dataset from \href{https://www.worldbank.org/ext/en/home}{worldbank.org} containing economic metrics for countries worldwide. Our goal is to predict Gross Domestic Product (GDP), a measure of the total value of goods and services produced within each country. We include Foreign Direct Investment (FDI) outflows (investments by a country's residents abroad), FDI inflows (foreign investments into a country), trade as a percentage of GDP, unemployment rate, and inflation measured the ratio of GDP in current local currency to GDP in constant local currency. The random effect will be modeled using regions of the world as determined by World Bank. 

We evaluate the MSPE of GTIMM, GLMER, LMM, RF, and a decision tree using data from 2022. Although the dataset initially contains 269 observations, we remove missing entries and dependencies, resulting in 97 countries. We then adjust the region categories to ensure adequate representation, settling on Africa, the Americas, Asia-Pacific, and Europe and Central Asia.

\begin{figure}[h!]
\centering
\scalebox{0.8}{
\begin{forest}
for tree={
  grow=south,
  parent anchor=south,
  child anchor=north,
  edge={-latex},
  rounded corners,
  draw,
  align=center,
  l sep+=15pt,
  s sep+=15pt
}
[
  {$\textbf{FDIOutflows} < 0.57$}
  [
    {$\textbf{FDIOutflows} < 0.015$}
    [
      {$\textbf{Trade} < -0.17$}
      [{\textbf{Node 1}\\[-4pt] -0.64 \\[-2pt] 38\%}]
      [{\textbf{Node 2}\\[-4pt] -0.13 \\ [-2pt] 23\%}]
    ]
    [{\textbf{Node 3}\\[-4pt] 0.26 \\[-2pt] 24\%}]
  ]
  [
    {\textbf{Node 4}\\[-4pt] 1.40 \\[-2pt] 15\%}
]
]
\end{forest}
}
\captionsetup{labelfont=bf}
\caption{\textbf{Decision Tree Splits}. The tree above displays the splits of the standardized data in a four-node decision tree used when modeling GTIMM. If the condition of the split is met, the data follows the left branch. The first value is the predicted standardized GDP whereas the second value represents the percentage of data in each node.}
\label{dafig1}
\end{figure}
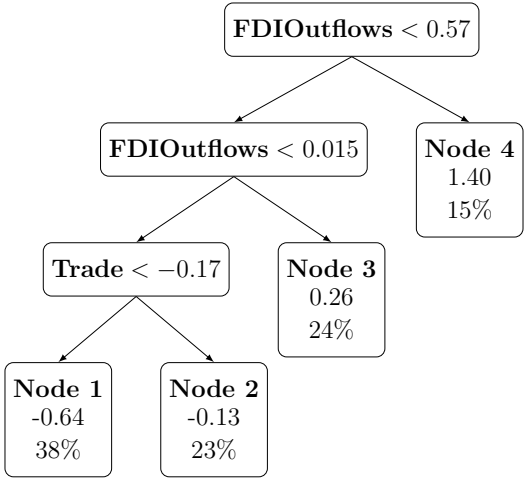

We use 5-fold cross-validation (CV) and find that the best tree has 4 terminal nodes with a learning rate of 0.01. We opt for 5-fold CV instead of the standard 10-fold CV to ensure a sufficient number of countries in each region appear in every fold, allowing the random effect to be modeled appropriately. \Cref{dafig1} shows the resulting tree splits: the first two splits use FDI Outflows predictor, while the third split uses Trade. Each terminal node contains at least 15\% of the dataset. 

\begin{figure}[h!] 
    \centering
    \includegraphics[scale=0.6]{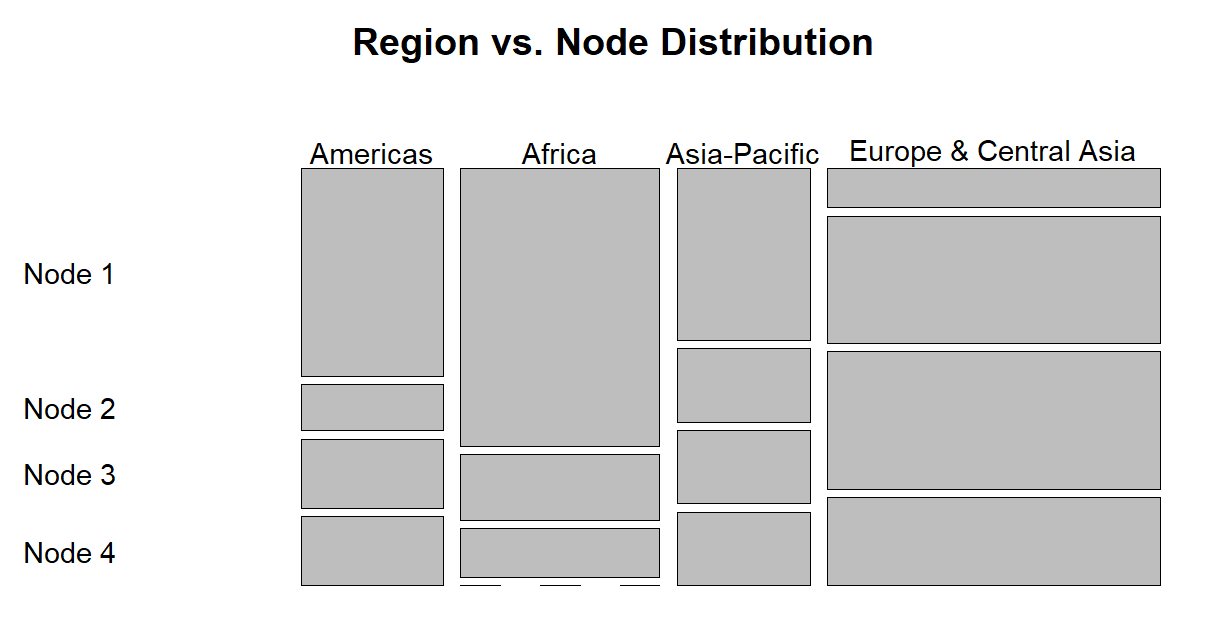}
    \captionsetup{labelfont=bf}
    \caption{\textbf{Region and Terminal Nodes Comparison} Nodes 1-4 refer to the terminal nodes of the tree from left to right. This mosaic diagram shows how the countries were distributed by both region classification and the terminal node of the decision tree. }
    \label{dafig2}
\end{figure}

\Cref{dafig2} illustrates how countries distribute across the terminal nodes and regions. Europe and Central Asia differs from the other three regions because most of its countries do not fall in the first node, while Africa exhibits the least diversity, with most of its countries in the first node, none in the second, and the rest in the third and fourth nodes. Important to note here that a linear mixed model would not capture this idea that for countries with the highest GDP (those in the 4th node), the random effect would have one less region. 

\begin{table}[h!]
    \centering
    \captionsetup{labelfont=bf}
    \caption{\textbf{Estimated GTIMM Coefficients for GDP Data.} Each column shows the fitted coefficients of a linear model for the partitioned data points in each terminal node of the decision tree.}
    \label{datab1}
    \small
    \begin{tabular}{lcccc}
        \toprule
        \textbf{Parameter} & \textbf{ Node 1} & \textbf{Node 2} & \textbf{Node 3} & \textbf{Node 4} \\
        \midrule
        Slope          & -0.6813  & -0.0903  &  0.2421  &  1.4392  \\
        FDI Inflows    &  0.0595 &  0.1698 & -0.7506 & -0.1137 \\
        FDI Outflows   &  -0.1747 & -0.1833  & 0.1137  & 0.0453  \\
        Inflation      & 0.0112 & -1.9613  &  0.0266  &  2.3389  \\
        Trade          & -0.1078  & -0.5621  & -0.4249  &  0.3245  \\
        Unemployment   &  0.0187 & -0.5919 & -0.0931 & -0.8549 \\
        \bottomrule
    \end{tabular}
\end{table}

\Cref{datab1} displays the estimated coefficients for each subgroup. The relationship between GDP and the predictors varies across nodes. For example, standardized FDI inflows may associate positively with standardized GDP for Node 1 and Node 2, where countries have a smaller GDP on average, however, this is not true for Node 3 and Node 4.   

\begin{table}[h]
    \centering
    \captionsetup{labelfont=bf}
    \caption{\textbf{MSPE Comparison of the model on GDP data.} Comparison of MSPE for the generalized tree-informed mixed model, the linear mixed model, random forest model, and a singular decision tree.}
    \label{datab2}
    \small
    \begin{tabular}{lccccc}
        \toprule
        \textbf{Model} & \textbf{GTIMM} & \textbf{LMM} & \textbf{RF} & \textbf{Tree} & \textbf{GLMER} \\ 
        \midrule
        \textbf{MSPE} & 0.385 & 0.729 & 0.618 & 0.631 & 0.722 \\
        \bottomrule
    \end{tabular}
\end{table}

\Cref{datab2} shows that the GTIMM model outperformed the other models, which were otherwise pretty similar in their MSPE. While the GLMER function in \textsf{R} can capture this tree-based mixed model regression, and does well in the simulation, with this dataset it used a tree with 1 terminal node, effectively fitting a linear mixed model. 

\section{Discussion}\label {discussion}
We have proposed a stochastic-gradient based estimation method using a quasi-likelihood framework to model a mixed effect regression tree. The GTIMM framework for modeling clustered data is characterized by both subgroup-specific patterns and overarching correlations. By combining the predictive capabilities of a decision tree and the interpretability of linear regressions, this is a powerful tool for modeling and understanding clustered data. Additionally, a theoretical bound for the MSPE of this estimation method has been found relative to the GLMM. Through simulation and data analysis, this model performs well to standard modeling techniques.

\renewcommand{\thesection}{\Alph{section}.\arabic{section}}
\appendix
\section{Appendix} \label{appendix}
\subsection{Proof of \Cref{thm1}}
\begin{proof}
The difference in MSPE is
\begin{align*}
    \begin{split}
         &\left| E \left[Y_{i} - g^{-1}\left(\hat{T}_{i}^{(m)} \right) \right]^2 - E \left[Y_{i} - g^{-1}\left(\hat{M}_{i} \right) \right]^2 \right| \\
         &= \left |E \left(h^2 (\hat{T}_{i}^{(m)}) - h^2(\hat{M}_{i}) \right) - 2E \left(Y_{i} \left(h(\hat{T}_{i}^{(m)}) - h(\hat{M}_{i}) \right) \right)\right| \\
         &\leq E \left|h(\hat{T}_{i}^{(m)}) - h(\hat{M}_{i}) \right| \left|h(\hat{T}_{i}^{(m)}) - h(\hat{M}_{i}) - 2Y_{i} \right| \\
         &= E \left|\hat{T}_{i}^{(m)} - \hat{M}_{i} \right| \left|\hat{T}_{i}^{(m)} - \hat{M}_{i} - 2Y_{i} \right|.
    \end{split}
\end{align*}
By H\"{o}lder's inequality,
\begin{align*}
    \begin{split}
         &\left| \text{MSPE}_{\widehat{\text{GTIMM}}} - \text{MSPE}_{\widehat{\text{GLMM}}} \right| \\
         &\leq E \left|\hat{T}_{i}^{(m)} - \hat{M}_{i} \right| \left|\hat{T}_{i}^{(m)} - \hat{M}_{i} - 2Y_{i} \right| \\
         &\leq \sqrt{E \left|\hat{T}_{i}^{(m)} - \hat{M}_{i} \right|^2 E\left|\hat{T}_{i}^{(m)} - \hat{M}_{i} - 2Y_{i} \right|^2}.
    \end{split}
\end{align*}
Consider the term
\begin{equation}
    \label{error_eq}
    \begin{split}
    \left|\hat{T}_{i}^{(m)} - \hat{M}_{i} \right|
    &= E \left|\bm{X}_i^\top \hat{\bm{\beta}}^* \bm{R}_i + \bm{Z}_i^\top \hat{\bm{b}} - \bm{X}_i^\top \hat{\bm{\beta}} + \bm{Z}_i^\top \tilde{\bm{b}} \right| \\
    &\leq \left|\bm{X}_i^\top \hat{\bm{\beta}}^* \bm{R}_i - \bm{X}_i^\top \hat{\bm{\beta}} \right| + \left|\bm{Z}_i^\top \hat{\bm{b}} - \bm{Z}_i^\top \tilde{\bm{b}} \right|,
    \end{split}
\end{equation} 
where $\hat{\bm{b}}$ and $\tilde{\bm{b}}$ are given by 
\begin{center}
    $\hat{\bm{b}} = \displaystyle \frac{\bm{\Sigma_{b}} \bm{Z}_i}{\left(\bm{\Sigma_{\varepsilon}} \right)_{ii} + \bm{Z}_i^{\top} \bm{\Sigma_{b}} \bm{Z}_i}  \left(Y_{i} - \bm{X}_i^\top \hat{\bm{\beta}}^* \bm{R}_i \right)$
\end{center}
\begin{center}
    $\tilde{\bm{b}} = \displaystyle \frac{\bm{\Sigma_{b}} \bm{Z}_i}{\left(\bm{\Sigma_{\varepsilon}} \right)_{ii} + \bm{Z}_i^{\top} \bm{\Sigma_{b}} \bm{Z}_i}  \left(Y_{i} - \bm{X}_i^\top \hat{\bm{\beta}} \right)$.  
\end{center}
Plugging-in $\hat{\bm{b}}$ and $\tilde{\bm{b}}$ in \Cref{error_eq}, one has
\begin{equation}
    \begin{split}
    \left|\hat{T}_{i}^{(m)} - \hat{M}_{i} \right| 
    &\leq \left|\bm{X}_i^\top \hat{\bm{\beta}}^* \bm{R}_i - \bm{X}_i^\top \hat{\bm{\beta}} \right| + \left| \displaystyle \frac{\bm{Z}_i^\top \bm{\Sigma_{b}} \bm{Z}_i}{\left(\bm{\Sigma_{\varepsilon}} \right)_{ii} + \bm{Z}_i^{\top} \bm{\Sigma_{b}} \bm{Z}_i} \right| \left|\bm{X}_i^\top \hat{\bm{\beta}}^* \bm{R}_i - \bm{X}_i^\top \hat{\bm{\beta}} \right| \\
    &= \left| 1 + \displaystyle \frac{\bm{Z}_i^\top \bm{\Sigma_{b}} \bm{Z}_i}{\left(\bm{\Sigma_{\varepsilon}} \right)_{ii} + \bm{Z}_i^{\top} \bm{\Sigma_{b}} \bm{Z}_i} \right| \left|\bm{X}_i^\top \hat{\bm{\beta}}^* \bm{R}_i - \bm{X}_i^\top \hat{\bm{\beta}} \right| \\
    &\leq 2 \left|\bm{X}_i^\top \hat{\bm{\beta}}^* \bm{R}_i - \bm{X}_i^\top \hat{\bm{\beta}} \right|.
    \end{split}
\end{equation}
So, 
\begin{align*}
    \begin{split}
    E \left|\hat{T}_{i}^{(m)} - \hat{M}_{i} \right|^2 
    &\leq 4 E \left|\bm{X}_i^\top \hat{\bm{\beta}}^* \bm{R}_i - \bm{X}_i^\top \bm{\beta}^* \bm{R}_i - \bm{X}_i^\top \hat{\bm{\beta}} + \bm{X}_i^\top \bm{\beta} + \bm{X}_i^\top \bm{\beta}^* \bm{R}_i - \bm{X}_i^\top \bm{\beta} \right|^2 \\
    &\leq 4 E \left(\left|\bm{X}_i^\top \hat{\bm{\beta}}^* \bm{R}_i - \bm{X}_i^\top \bm{\beta}^* \bm{R}_i\right| + \left|\bm{X}_i^\top \hat{\bm{\beta}} - \bm{X}_i^\top \bm{\beta} \right| + \left|\bm{X}_i^\top \bm{\beta}^* \bm{R}_i - \bm{X}_i^\top \bm{\beta} \right| \right)^2 \\
    &\leq 36 \max \left\{T_1, T_2, T_3 \right\},
    \end{split}
\end{align*}
where 
\begin{center}
    $T_1 = E\left(\bm{X}_i^\top \hat{\bm{\beta}}^* \bm{R}_i - \bm{X}_i^\top \bm{\beta}^* \bm{R}_i\right)^2$, \\

    $T_2 = E\left(\bm{X}_i^\top \hat{\bm{\beta}} - \bm{X}_i^\top \bm{\beta} \right)^2$, \\

    $T_3 = E\left(\bm{X}_i^\top \bm{\beta}^* \bm{R}_i - \bm{X}_i^\top \bm{\beta} \right)^2$.
\end{center}

Using the moment inequality from Proposition 3.2 from \cite{rivasplata2012subgaussian} on the sub-Gaussian random variable $\hat{T}_{i}^{(m)} - \hat{M}_{i} - 2Y_{i}$, whose second moment is upper bounded by 
\begin{center}
    $4 \tilde{\sigma}_i^2 = 4 \left[\text{Var}\left(Y_{i} - \hat{T}_{i}^{(m)} \right) + \text{Var}\left(Y_{i} - \hat{M}_{i} \right) \right]$, 
\end{center}
yields
\begin{center}
    $\left| \text{MSPE}_{\widehat{\text{GTIMM}}} - \text{MSPE}_{\widehat{\text{GLMM}}} \right| 
    \leq 12 \tilde{\sigma}_i \max \left\{T_1, T_2, T_3 \right\}$.
\end{center} 
The $T_1$ term represents the estimation error of the fixed effect in GTIMM for the $n_m$ observations in the $m$th region where the $i$th observation falls, so $T_1 = \mathcal{O}(\frac{M}{N})$. As $T_2$ term represents the estimation error of fixed effect in GLMM, we have $T_2 = \mathcal{O}(\frac{1}{N})$. Since $\bm{R}_i$ is a sparse indicator vector, and $M$ is small relative to $N$, the contribution of the tree structure to the overall error is negligible. In other words, the error due to the tree structure does not significantly impact the overall bound because $M$ is small. Therefore, $T_3 = \mathcal{O}\left(\frac{1}{N}\right)$. We get 
\begin{center}                              
    $\left|\text{MSPE}_{\widehat{\text{GTIMM}}} - \text{MSPE}_{\widehat{\text{GLMM}}} \right| = \mathcal{O}\left(\frac{M}{N}\right)$.
\end{center}
\end{proof}

\medskip
\bibliographystyle{apacite}

\bibliography{reference}
\end{document}